\documentclass[preprint,amsmath, amssymb, nofootinbib, aps, prd]{revtex4-1}

\usepackage{graphicx}
\usepackage{tikz}

\newcommand{\eps}{\epsilon}
\newcommand{\nn}{\nonumber}
\newcommand{\tr}{{\rm Tr}}

\newcommand{\overbar}[1]{\mkern 1.5mu\overline{\mkern-1.5mu#1\mkern-1.5mu}\mkern 1.5mu}
\renewcommand{\bar}{\overbar}
\renewcommand{\tilde}{\widetilde}

\newcommand{\cnab}[1]{
\begin{tikzpicture}[#1]
\draw (0,0)--(2,4)--(-2,4)--cycle;
\draw (0.5,1)--(-1,4);
\end{tikzpicture}
}
\newcommand{\varnab}{\cnab{scale=0.07}}

\newcommand{\cdel}[1]{
\begin{tikzpicture}[#1]
\draw (-2,0)--(2,0)--(0,4)--cycle;
\draw (-1,0)--(0.5,3);
\end{tikzpicture}
}
\newcommand{\vardel}{\cdel{scale=0.07}}

\begin{document}

\title{
${\cal N} = 2$ super Yang-Mills theory in Projective Superspace
}

\author{Ariunzul Davgadorj}
\email{ariunzul.d@gmail.com}

\author{Rikard von Unge}
\email{unge@physics.muni.cz}

\affiliation{Institute for Theoretical Physics, Masaryk University, \\  611 37 Brno, Czech Republic}
\date{\today}

\begin{abstract}
We find a formulation of $\mathcal{N}=2$ supersymmetric Yang-Mills theory in Projective superspace. In particular we find an expression for the field strength in terms of an unconstrained prepotential which is desirable when quantizing the theory. We use this to write the action in terms of the prepotential and show that it reduces to the known result in the abelian limit.
\end{abstract}

\maketitle
\tableofcontents

\section{Introduction}
Projective superspace \cite{Karlhede:1984vr} is a manifestly supersymmetric formalism for theories with eight supercharges. The fact that the Projective superspace formalism is closely connected with the Twistor space description of Hyperk\"ahler manifolds \cite{Hitchin:1986ea,Karlhede:1986mg,Lindstrom:2008gs} or Quaternion K\"ahler manifolds \cite{deWit:2001brd} has led to many applications in mathematics and physics. For instance, both the Hyperk\"ahler quotient construction \cite{Hitchin:1986ea} and the physical explanation of the wall crossing phenomenon \cite{Gaiotto:2008cd,Crichigno:2012vd} made use of projective superspace. Moreover, aspects of gauge theory has been developed \cite{Lindstrom:1989ne,GonzalezRey:1997db} as well as propagators and Feynman rules for matter multiplets \cite{GonzalezRey:1997qh,GonzalezRey:1997xp}. More recently, also supergravity has been treated in the Projective superspace formalism \cite{Kuzenko:2007cj,Kuzenko:2007hu,Kuzenko:2008wr,Kuzenko:2008ep,Kuzenko:2009zu}. For an introduction we refer to \cite{Kuzenko:2010bd}.

Alternatively one may use Harmonic superspace \cite{Galperin:1984av} which also provides a manifestly supersymmetric formalism for theories with eight supersymmetries. Harmonic superspace also represents a mature formalism with all the basic ingredients developed. In particular, nonabelian Yang-Mills theory was formulated already some time ago in \cite{Zupnik:1986ca,Zupnik:1987vm}. A nice introduction to the topic can be found in the book \cite{Galperin:2007wpa}.

In both the Projective and Harmonic formalism the superspace $\mathbb{R}^{4|8}$ is appended by the auxiliary factor $\mathbb{CP}^1$ where the R-symmetry group $SU(2)_R$ naturally acts and on which the superfields depend\footnote{This construction was originally suggested by Rosly in \cite{rosly1983super}.}. More precisely, in the Harmonic approach the superfields depend on the full $\mathbb{CP}^1$ factor and can be written in terms of the spinor harmonic basis of the $\mathbb{CP}^1 = S^2$ (hence the name). On the other hand, in the Projective case the superfields depend only holomorphically on the auxiliary $\mathbb{CP}^1$ factor. This difference might seem insignificant and it is interesting to observe that it leads to quite different properties of the two formalisms. The fact that in Harmonic superspace the full superspace integral is over the full volume of the auxiliary $S^2$ whereas in the Projective approach we get contour integrals over closed contours on $\mathbb{CP}^1$ leads to very different calculational techniques and details of the calculations. For instance, the close connection between the Laurent coefficients of the $\mathbb{CP}^1$ dependence of the superfield in Projective superspace and the ${\cal N}=1$ components makes the Projective superspace formalism ideal for discussing ${\cal N}=2$ supersymmetric sigma models. On the other hand, until recently, the Harmonic superspace formalism has been the only choice when doing quantum computations in Yang-Mills theory.
A careful comparison between the Harmonic and the Projective approach can be found in \cite{Kuzenko:1998xm,Butter:2012ta}.

Recently a hybrid formalism between the Projective and Harmonic superspaces has been constructed \cite{Jain:2009aj,Jain:2010gm,Jain:2012jx,Jain:2012zx,Jain:2013hua} called Hyperspace. Using these new tools the authors were able to derive several new results for Projective superspace, in particular for nonabelian Yang-Mills theory formulated in Projective superspace.

In this paper we rederive and extend these results based on a purely Projective superspace formalism. In particular we derive an expression for the Yang-Mills field strength in terms of the gauge prepotential superfield and show that it has the correct properties. We explicitly show that this Yang-Mills action agrees with the action derived from a one-loop computation with a hypermultiplet running in the loop. We furthermore show that in the abelian limit, our expressions reduce to known expressions in Projective superspace.

The paper is organized as follows: In the next section we give our conventions and review the basic facts about Yang-Mills theory in Projective superspace.
In section III we express the connection coefficients, and hence the field strength, in terms of the gauge prepotential and in section IV we show that these objects transform correctly. Using these results, in section V we prove that the one-loop form of the action agrees with the action constructed using the field strength superfield and finally in section VI we show that our construction correctly reduces to known results in the abelian limit. In three appendices we review known results that nevertheless are crucial for this paper; we show how from extracting the divergent part of a one loop calculation with a hypermultiplet coupled to a background gauge field, we can extract a closed form for the Yang-Mills action expressed in terms of a projective prepotential. In appendix B it is shown that the derived action is gauge invariant. This form of the action is used in the paper as a check on our expression for the field strength. In the final appendix we discuss aspects of the $\epsilon$-prescription introduced in \cite{Jain:2009aj} which is an integral part of the techniques used to prove the results of this article.

\section{Setup and conventions}
The ${\cal N}=2$ algebra of super covariant derivatives is given by
\begin{align}
\{D^a_\alpha,D^b_\beta\} = 0 \;\; , \;\;\;\{D^a_\alpha , \bar{D}_{b\dot\alpha} \} = \delta^a_b i\partial_{\alpha\dot\alpha}\;\;.
\end{align}
In the following we will use the notation $D \equiv D^1$ and $Q \equiv D^2$.
In order to be able to construct actions without higher derivatives we need to define an anticommuting subalgebra of four supercovariant derivatives which can then be used to constrain the physical superfields. We do this by introducing a coordinate on an auxiliary $\mathbb{CP}^1$ manifold $\zeta$ and using this coordinate we construct the supercovariant derivatives
\begin{align}
\nabla_\alpha = D_\alpha + \zeta Q_\alpha \;\; , & &
\bar \nabla _{\dot\alpha} = \bar Q_{\dot\alpha} - \zeta \bar D_{\dot\alpha}\;\;,
\end{align}
{}from which follows that $\{\nabla_\alpha,\nabla_\beta\}=0=\{\nabla_\alpha,\bar\nabla_{\dot\alpha}\}$. We may furthermore define a second, linearly independent, set of supercovariant derivatives
\begin{align}
\Delta_{\alpha} = Q_\alpha-\frac1\zeta D_\alpha\;\;, & &
\bar\Delta_{\dot\alpha} = \bar D_{\alpha}+\frac1\zeta \bar Q_{\dot\alpha}\;\;,
\end{align}
which also anticommute among themselves. The only non-zero anti commutators are
\begin{align}
\{\nabla_\alpha,\bar\Delta_{\dot\alpha}\}=\{\Delta_{\alpha},\bar\nabla_{\dot\alpha}\} = 2i\partial_{\alpha\dot\alpha}.
\end{align}
Physical superfields are annihilated by all the $\nabla$ and $\bar\nabla$ derivatives; we call such superfields {\em projective} superfields. By choosing the $\zeta$ dependence of the superfields we get different representations.

Among the matter multiplets we define the ${\cal O}(k)$ multiplets whose $\zeta$ dependence is given by the fact that they transform as section of the ${\cal O}(k)$ complex line bundle on $\mathbb{CP}^1$.

We also define the Polar multiplet $(\Upsilon,\bar\Upsilon)$ where $\Upsilon(\zeta)$ is analytic around the north pole and hence called ``arctic'' while $\bar\Upsilon(\zeta)$ is analytic around the south pole and thus called ``antarctic''.

Finally we also consider superfields with $\zeta$ dependence singular both at the north and south pole which we accordingly call ``tropical''.

Conjugation in projective superspace is defined by combining Hermitian conjugation with the antipodal map on $\mathbb{CP}^1$
\begin{align}
\bar{f(\zeta)} = \zeta^{p} f^*(-\frac1\zeta)\;\;,
\end{align}
where the number $p$ is determined by the representation.

\subsection{Measures}
The full ${\cal N}=2$ superspace measure is constructed by an integral over all Grassmann coordinates $d^8\theta = D^2Q^2\bar{D}^2\bar{Q}^2$ as well as an integral over space $d^4 x$ and a contour integral on $\mathbb{CP}^1$. In this paper the contour integral measure will always have a factor of $2\pi i$ in the denominator which we, to avoid cluttering, will suppress in all subsequent formulas
\begin{align}
\oint \frac{d\zeta}{2\pi i} \rightarrow \oint d\zeta \;\;\;.
\end{align}

To avoid higher derivatives, physical actions will be constructed using Projective superfields and consequently integrated over only half of the Grassmann coordinates. Since Projective superfields are annihilated by $\nabla,\bar\nabla$ we define the measure (up to a $\zeta$ dependent factor) as $\Delta^2\bar\Delta^2 = \Delta^4$. However, since
\begin{align}
\Delta_\alpha = \frac1\zeta \nabla_\alpha - \frac2\zeta D_\alpha
\;\;, &&\bar\Delta_\alpha = \frac1\zeta\bar{\nabla}_\alpha+2\bar{D}_\alpha\;\;,
\end{align}
we may always write $\Delta^4 \propto D^2\bar{D}^2$ when acting on a projective Lagrangian.
Therefore we will define the Projective measure as
\begin{align}
\oint d\zeta\int d^4 x \int [d^4\theta]_P   = \oint d\zeta\int d^4 x D^2\bar{D}^2 \;\; .
\end{align}
For instance, the action for a Polar hypermultiplet is given by
\begin{align}
\int d^4 x \int [d^4\theta]_P \oint \frac{d\zeta}{\zeta} \bar\Upsilon \Upsilon \;\; .
\end{align}

Finally we will also deal with ${\cal N}=2$ chiral fields. A chiral field $F$ satisfies the constraint
$\bar{D}_{\dot\alpha} {F} = \bar{Q}_{\dot\alpha}{F} = 0$ which, using the projective supercovariant derivatives, can be written $\bar\nabla_{\dot\alpha} {F} = \bar\Delta_{\dot\alpha} {F} = 0$. We may construct ${\cal N}=2$ superpotential terms using the ${\cal N}=2$ chiral measure as
\begin{align}
\oint d\zeta\int d^4 x \int [d^4\theta]_C  f(F) =  \oint d\zeta\int d^4 x D^2Q^2 f(F)\;\;.
\end{align}
The chiral field we will deal with in this article is the ${\cal N}=2$ field strength $\mathbb{W}$ which however is ${\cal N}=2$ gauge covariantly chiral. In that case, the action for super Yang-Mills theory can be written as
\begin{align}\label{eq:YMA}
\frac12\int d^4 x \int [d^4\theta]_C \oint \frac{d\zeta}{\zeta}\tr(\mathbb{WW})\;\;\;,
\end{align}
where the measure is constructed using the gauge supercovariant derivatives and the $\zeta$ integral is trivial since $\mathbb{W}$ is $\zeta$ independent.

\subsection{Yang-Mills theory in Projective Superspace}
The ${\cal N}=2$ algebra of super gauge covariant derivatives is given by\footnote{Many results in this section have appeared before in \cite{Lindstrom:1989ne,GonzalezRey:1997db,Kuzenko:2014mva,Butter:2014xxa}}
\begin{align}
\{{\mathbb D}_\alpha,{\mathbb Q}_\beta\} = iC_{\alpha\beta}\bar{\mathbb W} \;\; , \;\;\{{\mathbb D}_\alpha , \bar{\mathbb D}_{\dot\alpha} \} = i {\cal\varnab}_{\alpha\dot\alpha} \;\; , \;\;
\{{\mathbb Q}_\alpha , \bar{\mathbb Q}_{\dot\alpha} \} = i {\varnab}_{\alpha\dot\alpha} \;\;,
\end{align}
where $\bar{\mathbb{W}}$ is the ${\cal N}=2$ field strength. A Bianchi identity shows it is antichiral $\mathbb{D}_{\alpha} \bar{\mathbb{W}} = \mathbb{Q}_{\alpha} \bar{\mathbb{W}} = 0$.
The gauge covariant projective superspace derivatives are constructed as
\begin{align}
\varnab_\alpha = {\mathbb D}_\alpha + \zeta {\mathbb Q}_\alpha\;\;,& &
\bar \varnab _{\dot\alpha} = \bar{\mathbb Q}_{\dot\alpha} - \zeta \bar {\mathbb D}_{\dot\alpha}\;\;, \\
\vardel_{\alpha} = {\mathbb Q}_\alpha-\frac1\zeta {\mathbb D}_\alpha\;\;, & &
\bar\vardel_{\dot\alpha} = \bar {\mathbb D}_{\alpha}+\frac1\zeta \bar{\mathbb Q}_{\dot\alpha}\;\; ,
\nn
\end{align}
and it is interesting to notice that the $\varnab$ derivatives still anticommute among themselves. The field strength appears only if we anticommute derivatives at different $\zeta$ coordinates
\begin{align}
\{\varnab_\alpha(\zeta_1),\varnab_{\beta}(\zeta_2)\} = i(\zeta_2-\zeta_1)C_{\alpha\beta}\bar{\mathbb W}\;\; ,
\end{align}
or by introducing the $\zeta$ derivative $\partial_\zeta$ as
\begin{align}\label{W}
\{\varnab_\alpha , [\partial_\zeta,\varnab_\beta]\} = iC_{\alpha\beta}\bar{\mathbb W} \;\; .
\end{align}
By construction $\varnab_\alpha\bar{\mathbb W} = 0$. Furthermore, using (\ref{W}) there is a Bianchi identity that relates $\mathbb W$ and $\bar{\mathbb W}$ as $\varnab^2{\mathbb W} = \bar\varnab^2\bar{\mathbb W}$.

The polar multiplets can be made to transform under a gauge transformation as
\begin{align}
\Upsilon \rightarrow e^{i\Lambda}\Upsilon \;\;,\\
\bar\Upsilon \rightarrow \bar\Upsilon e^{-i\bar\Lambda} \;\;.\nn
\end{align}
The gauge parameters are ${\cal N}=2$ superfields but need to be constrained to respect the constraints of the transforming field. Thus $\Lambda$ is arctic projective and $\bar\Lambda$ is antarctic projective. Precisely in analogy with ${\cal N}=1$ super Yang-Mills theory, to make an invariant action we introduce a real projective superfield $V$ that converts $\Lambda$ gauge transformations to $\bar\Lambda$ gauge transformations (and the reverse). The simply transforming object is $e^V$ with the transformation properties
\begin{align}
e^{V} \rightarrow e^{i\bar\Lambda}e^Ve^{-i\Lambda} \;\; .
\end{align}
From the transformation it is clear that $V$ has to have tropical dependence on $\zeta$.

Using $e^V$ as a converter we can make all fields transform with only the arctic $\Lambda$-parameter
\begin{align}
\tilde{\Upsilon} &\equiv \Upsilon \rightarrow e^{i\Lambda}\tilde{\Upsilon}\;\; , \nn\\
 \bar{\tilde{\Upsilon}} &\equiv \bar\Upsilon e^V \rightarrow \bar{\tilde{\Upsilon}}e^{-i\Lambda} \;\; ,
\end{align}
or with the antarctic $\bar\Lambda$-parameter
\begin{align}
\tilde{\Upsilon} &\equiv e^V\Upsilon \rightarrow e^{i\bar\Lambda}\tilde{\Upsilon} \;\;, \nn\\
 \bar{\tilde{\Upsilon}} &\equiv \bar\Upsilon \rightarrow \bar{\tilde{\Upsilon}}e^{-i\bar\Lambda} \;\; .
\end{align}
These choices will be called the gauge arctic or gauge antarctic representation respectively. Note that both $\nabla$ and $\bar\nabla$ are covariant in the arctic and antarctic representations without any additional connection coefficient since both $\Lambda$ and $\bar\Lambda$ are projective.

We may finally define the gauge vector representation where the fields transform with a real, $\zeta$-independent superfield $K$. We achieve this by splitting $e^V$ into its arctic and antarctic parts
\begin{align}
e^V = e^{\bar U}e^U \;\; ,
\end{align}
where $U$ is arctic and $\bar U$ is antarctic while neither of them are projective, and transforming under supergauge transformations as
\begin{align}
e^U &\rightarrow e^{iK} e^U e^{-i\Lambda} \;\; , \nn \\
e^{\bar U} &\rightarrow e^{i\bar\Lambda} e^{\bar U} e^{-iK} \;\; ,
\end{align}
and then defining the covariantly transforming superfields as
\begin{align}
\tilde{\Upsilon} &\equiv e^{U}\Upsilon \rightarrow e^{i K}\tilde{\Upsilon} \;\;, \nn\\
\bar{\tilde{\Upsilon}} &\equiv \bar\Upsilon e^{\bar U} \rightarrow \bar{\tilde{\Upsilon}}e^{-iK} \;\;.
\end{align}
They are gauge covariantly projective since they are annihilated by the gauge covariant derivatives
\begin{align}\label{GCov}
\varnab_\alpha
&= e^U \nabla_\alpha e^{-U} = e^{-\bar U} \nabla_\alpha e^{\bar U} \;\;, \\
\overbar{\varnab}_{\dot\alpha}
&= e^U \overbar{\nabla}_{\dot\alpha} e^{-U} = e^{-\bar U} \overbar{\nabla}_{\dot\alpha} e^{\bar U} \;\;. \nn
\end{align}
The fact that $\varnab_\alpha$ $(\overbar{\varnab}_{\dot\alpha})$ can be written purely in terms of either $U$ or $\bar{U}$ follows from $V$ being projective which allows us to write
\begin{align}
0= \nabla_\alpha e^V = (\nabla_\alpha e^{\bar U}) e^{U} + e^{\bar U}(\nabla_\alpha e^{U}) \;\;,
\end{align}
which when multiplied by $e^{-\bar U}$ from the left and $e^{-U}$ from the right gives (\ref{GCov}) and allows us to identify the spinorial gauge connection
\begin{align}
\Gamma_\alpha(\zeta) = e^U (\nabla_\alpha e^{-U}) = e^{-\bar U} (\nabla_\alpha e^{\bar U}) \;\;.
\end{align}
Since $U$ is arctic, from the first relation we deduce that $\Gamma_\alpha(\zeta)$ contains no negative powers of $\zeta$. At the same time, since $\bar{U}$ is antarctic, the second relation tells us that $\Gamma_\alpha(\zeta)$ contains only negative powers of $\zeta$ as well as a constant term and a linear term. We conclude that $\Gamma_\alpha(\zeta)$ only has a constant and linear term $\Gamma_\alpha = \Gamma^1_\alpha + \zeta{\Gamma}^2_\alpha$ so that
\begin{align}
{\mathbb D}_\alpha &= D_\alpha + {\Gamma}^1_\alpha \;\; ,\nn \\
{\mathbb Q}_\alpha &= Q_\alpha + {\Gamma}^2_\alpha \;\;.
\end{align}
The field strength can be written
\begin{align}
\bar{\mathbb{W}} = \frac i2 \left(D^\alpha\Gamma^2_\alpha -Q^\alpha\Gamma^1_\alpha
+\{\Gamma^{1\alpha},\Gamma^2_\alpha\}\right) \;\;.
\end{align}

To find the field strength in the arctic or antarctic representations we start with relation (\ref{W}) and conjugate it with either $e^U$ to get the arctic representation or $e^{\bar U}$ to get the antarctic representation
\begin{align}\label{aWdef}
\{\nabla_\alpha ,[e^{-U}\partial_\zeta e^{U},\nabla_\beta]\} = i C_{\alpha\beta} \;e^{-U}\bar {\mathbb W} e^{U}\;\;, \nn\\
\{\nabla_\alpha ,[e^{\bar U}\partial_\zeta e^{-\bar U},\nabla_\beta]\} = i C_{\alpha\beta} \;e^{\bar U}\bar{\mathbb W}
e^{-\bar U}\;\;.
\end{align}
In the arctic representation we thus define the gauge covariant $\zeta$-derivative and the field strength as
\begin{align}
{\cal D}_\zeta &= \partial_\zeta + A_\zeta = e^{-U}\partial_\zeta e^U\;\;, \nn\\
\bar{\cal W}(\zeta) &= e^{-U} \bar{\mathbb W} e^{U} \;\;,
\end{align}
whereas for the antarctic representation we define
\begin{align}
\widetilde{\cal D}_\zeta &= \partial_\zeta + \widetilde A_\zeta = e^{\bar U}\partial_\zeta e^{-\bar U}\;\;, \nn\\
\bar{\widetilde{\cal W}}(\zeta) &= e^{\bar U} \bar{\mathbb W} e^{-\bar U}\;\;.
\end{align}
From (\ref{aWdef}) we find that
\begin{align}
\bar{\cal W} &= -i \nabla^2 A_\zeta \;\;,\nn \\
\bar{\widetilde{\cal W}} &= -i \nabla^2 \widetilde{A}_\zeta\;\;,
\end{align}
and similarly
\begin{align}
{\cal W} &= -i \bar{\nabla}^2 A_\zeta \;\;,\nn \\
{\widetilde{\cal W}} &= -i \bar\nabla^2 \widetilde{A}_\zeta\;\;.
\end{align}
We notice that the constraints and Bianchi identity derived from (\ref{W}) and rewritten in the arctic (antarctic) representation are automatically satisfied by these expressions.

Notice that $\bar{\cal W}(\zeta)$ depends arctically on $\zeta$ whereas $\bar{\tilde{\cal W}}$ depends antarctically. Furthermore $A_\zeta$ contains all positive powers of $\zeta$ as well as a constant term while $\tilde{A}_\zeta$ contains powers of $\zeta$ smaller or equal than $-2$.

Finally we notice that there is a relation between $A_\zeta$ and $\widetilde{A}_\zeta$ involving only $e^V$ since
\begin{align}\label{useful}
e^{-V}(\partial_\zeta e^V) = e^{-U}(\partial_\zeta e^{U}) + e^{-U}e^{-\bar U}(\partial_\zeta e^{\bar U}) e^{U} =
A_\zeta - e^{-V}\tilde{A}_\zeta e^{V}\;\; .
\end{align}
In the rest of the paper we will be working mainly in the arctic representation.

\section{The gauge potential $A_\zeta$}
What we really would like to do is to express $A_\zeta$ in terms of $V$. One way to do this would be to use (\ref{useful}) to write
\begin{align}
\partial_\zeta (e^V-1) = A_\zeta - \tilde{A}_\zeta + (e^V-1)A_\zeta - \tilde{A}_\zeta (e^V-1)\;\;.
\end{align}
We may now solve this equation recursively in powers of $X \equiv (e^V-1)$. We introduce the notation that $A_\zeta^{(n)}$ is the part of $A_\zeta$ proportional to $n$ powers of $X$. Then we get a recursion relation
\begin{align}\label{Arec}
A^{(1)}_\zeta - \tilde{A}^{(1)}_\zeta &= \partial_\zeta X \;\;,\\
A^{(n+1)}_\zeta - \tilde{A}^{(n+1)}_\zeta &= -XA^{(n)}_\zeta + \tilde{A}^{(n)}_\zeta X\;\;n\geq 1 \;\;. \nn
\end{align}
To find the individual $A^{(n)}_\zeta$ or $\tilde{A}^{(n)}_\zeta$ one has to project on positive or negative powers of $\zeta$. Introducing the operators $\Pi_{\pm}(\cdot)$ that project onto positive (including the constant term) or negative powers of $\zeta$, this leads to the following solution for $A_\zeta$
\begin{align}\label{AsolP}
A^{(1)}_\zeta &= \Pi_{+} (\partial_\zeta X) \;\;,
\\ \nn
A^{(2)}_\zeta &= -\Pi_{+}(X \Pi_{+}(\partial_\zeta X) + \Pi_{-}(\partial_\zeta X) X)\;\;,
\\ \nn
A^{(3)}_\zeta &= \Pi_{+} \left[ X\Pi_{+}(X \Pi_{+}(\partial_\zeta X)) + \Pi_{-}(X \Pi_{+}(\partial_\zeta X)) X\right.
\\ \nn
& \left.+X\Pi_{+}(\Pi_{-}(\partial_\zeta X) X)) + \Pi_{-}(\Pi_{-}(\partial_\zeta X) X) X\right]\;\;,
\\ \nn
A^{(4)}_\zeta &= \ldots \;\;,\nn
\end{align}
and for $\tilde{A}_\zeta$
\begin{align}
\tilde{A}^{(1)}_\zeta &= -\Pi_{-}(\partial_\zeta X) \;\;,\\ \nn
\tilde{A}^{(2)}_\zeta &= \Pi_{-}(X \Pi_{+}(\partial_\zeta X) + \Pi_{-}(\partial_\zeta X) X) \;\;, \\ \nn
\tilde{A}^{(3)}_\zeta &= -\Pi_{-} [ X\Pi_{+}(X \Pi_{+}(\partial_\zeta X)) + \Pi_{-}(X \Pi_{+}(\partial_\zeta X)) X \\ \nn
& \left.+X\Pi_{+}(\Pi_{-}(\partial_\zeta X) X)) + \Pi_{-}(\Pi_{-}(\partial_\zeta X) X) X\right]\;\;, \\ \nn
\tilde{A}^{(4)}_\zeta &= \ldots \;\;.\nn
\end{align}
From the recursion relation it is easy to see that $A^{(n)}_\zeta - \tilde{A}^{(n)}_\zeta$ will have an overall $(-1)^{n+1}$ sign and will consist of a sum of all possible terms starting with the $\partial_\zeta X$ which is then projected either on the positive or negative powers of $\zeta$ followed by a multiplication with $X$ from the left if the previous projection was on positive powers but from the right if the projection was on negative powers. This procedure is continued until we reach $n$ factors of $X$.

Using the $\epsilon$ prescription introduced in \cite{Jain:2009aj} and described in the appendix we can express the projection operators in terms of contour integrals. In particular we have
\begin{align}
\Pi_+(\partial_\zeta X)(\zeta_0) &= \int d\zeta_1 \frac{X_1}{\zeta_{10}^2}\;\;,\\
\Pi_-(\partial_\zeta X)(\zeta_0) &= -\int d\zeta_1 \frac{X_1}{\zeta_{01}^2}\;\;,
\end{align}
we may write the first term in the solution as
\begin{align}
A^{(1)}_\zeta &= \int d\zeta_1 \frac{X_1}{\zeta_{10}^2}\;\;, \\
\tilde{A}_\zeta^{(1)} &= \int d\zeta_1 \frac{X_1}{\zeta_{01}^2}\;\;.
\end{align}
For arbitrary $n$ we show that the solution to the recursion relation is
\begin{align}\label{AsolX}
A^{(n)}_\zeta(\zeta_0) &= (-1)^{n+1} \int d\zeta_1 \ldots \int d\zeta_n \frac{X_1\ldots X_n}
{\zeta_{10}\zeta_{21}\ldots\zeta_{n,n-1}\zeta_{n0}}\;\;, \\
\tilde{A}_\zeta^{(n)}(\zeta_0) &= (-1)^{n+1} \int d\zeta_1 \ldots \int d\zeta_n \frac{X_1\ldots X_n}
{\zeta_{01}\zeta_{21}\ldots\zeta_{n,n-1}\zeta_{0n}}\;\;,
\end{align}
by assuming this to be true and then showing that, when inserted into (\ref{Arec}), this implies the same relation for $A^{(n+1)}_\zeta$. We have
\begin{align}
A^{(n+1)}_\zeta(\zeta_0) = (-1)^{n+1} \int d\zeta_1 \frac{1}{\zeta_{10}} \left[
-X_1 \int d\zeta_2 \ldots \int d\zeta_{n+1} \frac{X_2\ldots X_{n+1}}
{\zeta_{21}\zeta_{32}\ldots\zeta_{n+1,n}\zeta_{n+1,1}}
\nn\right.\\ \left.
 +\int d\zeta_2 \ldots \int d\zeta_{n+1} \frac{X_2\ldots X_{n+1}}
{\zeta_{12}\zeta_{32}\ldots\zeta_{n+1,n}\zeta_{1,n+1}} X_1
\right]\;\;.
\end{align}
After relabeling the $\zeta$ coordinates we may write this as
\begin{align}
A^{(n+1)}_\zeta(\zeta_0) = (-1)^{n+1}
\int d\zeta_1 \ldots \int d\zeta_{n+1}
\frac{X_1\ldots X_{n+1}}{\zeta_{21}\ldots\zeta_{n+1,n}}
\left[ -\frac{1}{\zeta_{10}\zeta_{n+1,1}} + \frac{1}{\zeta_{n+1,1}\zeta_{n+1,0}}
\right]\;\;,
\end{align}
which, using the $\zeta$ identity (\ref{tri}), can be rewritten as
\begin{align}
A^{(n+1)}_\zeta(\zeta_0) = (-1)^{n+2}
\int d\zeta_1 \ldots \int d\zeta_{n+1}
\frac{X_1\ldots X_{n+1}}{\zeta_{21}\ldots\zeta_{n+1,n}}\frac{1}{\zeta_{10}\zeta_{n+1,0}}\;\;,
\end{align}
which indeed is the correct form.

Notice that the solution for $A_\zeta$ is very close but not identical to the form conjectured in \cite{Jain:2012zx}. There the authors remarked that their solution did not transform correctly as a connection. We now proceed to show that our solution does transform as expected.

\section{Transformation properties of $A_\zeta$ under gauge transformations}
We prove that our solution for $A_\zeta$ transforms correctly under infinitesimal gauge transformations
\begin{align}
A_\zeta &\rightarrow -i\partial_{\zeta}\Lambda +[i\Lambda , A_\zeta]\;\;,\nn\\
\tilde{A}_\zeta &\rightarrow -i\partial_{\zeta}\bar\Lambda + [i\bar\Lambda,\tilde{A}_\zeta]\;\;.
\end{align}
We will organize the proof using induction in powers of $X$. Under infinitesimal gauge transformations $X$ transforms as
\begin{align}
X \rightarrow i\bar\Lambda - i\Lambda +i\bar\Lambda X - X i\Lambda\;\;.
\end{align}
From the explicit solution (\ref{AsolP}) we see that
\begin{align}
\delta A^{(1)}_\zeta &= \Pi_+(i\partial_\zeta\bar\Lambda - i\partial_\zeta\Lambda+\partial_\zeta(i\bar\Lambda X - Xi\Lambda)) =
-i\partial_\zeta\Lambda + \Pi_+(\partial_\zeta(i\bar\Lambda X) - \partial_\zeta(X i\Lambda))\;\;,\nn\\
\delta \tilde{A}^{(1)}_\zeta &= -\Pi_-(i\partial_\zeta\bar\Lambda - i\partial_\zeta\Lambda+\partial_\zeta(i\bar\Lambda X - Xi\Lambda))
= -i\partial_\zeta\bar\Lambda - \Pi_-(\partial_\zeta(i\bar\Lambda X - Xi\Lambda))\;\;,\\
\delta A^{(2)}_\zeta &= -\Pi_+[(i\bar\Lambda-i\Lambda)A^{(1)}-\tilde{A}_\zeta^{(1)}(i\bar\Lambda-i\Lambda)
-X i\partial_\zeta\Lambda+i\partial_\zeta\bar\Lambda X] + {\mathcal O}(X^2) \;\;,\nn \\
\delta \tilde{A}^{(2)}_\zeta &= \Pi_-[(i\bar\Lambda-i\Lambda)A^{(1)}-\tilde{A}_\zeta^{(1)}(i\bar\Lambda-i\Lambda)
-X i\partial_\zeta\Lambda+i\partial_\zeta\bar\Lambda X] + {\mathcal O}(X^2)\;\;,\nn
\end{align}
where we have concentrated on terms independent of or linear in $X$.
It is clear that the $X$ independent part of $\delta A^{(1)}_\zeta$ and $\delta \tilde{A}^{(1)}_\zeta$ in (\ref{Arec}) gives the $-i\partial_\zeta\Lambda$ and $-i\partial_\zeta\bar\Lambda$ term in the transformation. Furthermore, when we add terms linearly dependent on $X$ we find
\begin{align}
\delta A^{(2)}_\zeta + \delta A^{(1)}_\zeta & = [i\Lambda,A^{(1)}_\zeta] +{\mathcal O}(X^2)\;\;,\nn\\
\delta \tilde{A}^{(2)}_\zeta + \delta \tilde{A}^{(1)}_\zeta & = [i\Lambda,A^{(1)}_\zeta] +{\mathcal O}(X^2)\;\;.
\end{align}

Using the notation that $[\ldots]^{(n)}$ picks out terms with precisely $n$ powers of $X$ we now assume that
\begin{align}\label{IndGA}
\left[\delta A^{(n)}_\zeta + \delta A^{(n-1)}_\zeta\right]^{(n-1)} &= [i\Lambda , A^{(n-1)}_\zeta]\;\;,\\
\left[\delta \tilde{A}^{(n)}_\zeta + \delta \tilde{A}^{(n-1)}_\zeta\right]^{(n-1)} &= [i\bar\Lambda, \tilde{A}^{(n-1)}_\zeta]
\;\;,
\end{align}
which we just showed is true for $n=2$.
From the recursion relation for $A^{(n)}_\zeta$ (\ref{Arec}) we have
\begin{align}
\delta A^{(n)}_\zeta -\delta \tilde{A}^{(n)}_\zeta = -\delta X A^{(n-1)}_\zeta - X \delta A^{(n-1)}_\zeta
+ \delta\tilde{A}^{(n-1)}_\zeta X + \tilde{A}^{(n-1)}_\zeta \delta X\;\;,
\end{align}
which gives us
\begin{align}
\left[\delta A^{(n)}_\zeta -\delta \tilde{A}^{(n)}_\zeta\right]^{(n)} =& -(i\bar\Lambda X - X i\Lambda)A^{(n-1)}_\zeta +
\tilde{A}^{(n-1)}_\zeta (i\bar\Lambda X - X i\Lambda) +\nonumber\\
 & \left[-X\delta A^{(n-1)}_\zeta +\delta\tilde{A}^{(n-1)}_\zeta X\right]^{(n)}\;\;,\\{}
\left[\delta A^{(n+1)}_\zeta -\delta \tilde{A}^{(n+1)}_\zeta\right]^{(n)} =&
-(i\bar\Lambda  -  i\Lambda) A^{(n)}_\zeta + \tilde{A}^{(n)}_\zeta(i\bar\Lambda - i\Lambda) +
\nonumber\\{}
 & \left[-X\delta A^{(n)}_\zeta +\delta\tilde{A}^{(n)}_\zeta X\right]^{(n)}\;\;.
\end{align}
This leads to
\begin{align}
\left[\delta A^{(n+1)}_\zeta + \right. & \left. \delta A^{(n)}_\zeta -
\delta \tilde{A}^{(n+1)}_\zeta - \delta \tilde{A}^{(n)}_\zeta\right]^{(n)} = \nn \\
& = -(i\bar\Lambda X - X i\Lambda)A^{(n-1)}_\zeta +
\tilde{A}^{(n-1)}_\zeta (i\bar\Lambda X - X i\Lambda) -\\
& (i\bar\Lambda  -  i\Lambda) A^{(n)}_\zeta + \tilde{A}^{(n)}_\zeta(i\bar\Lambda - i\Lambda) -
\nn \\ &
X\left[\delta A^{(n)}_\zeta + \delta A^{(n-1)}_\zeta\right]^{(n)}
+ \left[\delta \tilde A^{(n)}_\zeta + \delta\tilde A^{(n-1)}_\zeta\right]^{(n)}X \;\;,\nn
\end{align}
which, after using the assumption (\ref{IndGA}) can be written as
\begin{align}
-(i\bar\Lambda X - & X i\Lambda)A^{(n-1)}_\zeta +
\tilde{A}^{(n-1)}_\zeta (i\bar\Lambda X - X i\Lambda)- \nn \\ &
(i\bar\Lambda  -  i\Lambda) A^{(n)}_\zeta + \tilde{A}^{(n)}_\zeta(i\bar\Lambda - i\Lambda)
-X[i\Lambda,A^{(n-1)}_\zeta] +[i\bar\Lambda,\tilde A^{(n-1)}_\zeta]X
\\ & =
\{XA^{(n-1)}_\zeta - \tilde A^{(n-1)}_\zeta X - \tilde A^{(n)}_\zeta \} i\Lambda
+ i\Lambda\{-XA^{(n-1)}_\zeta+\tilde A^{(n-1)}_\zeta X - A^{(n)}_\zeta\} +\nn\\
& i\Lambda A^{(n)}_\zeta + \tilde A^{(n)}_\zeta i\bar\Lambda \;\;,\nn
\end{align}
and after using the $A_\zeta^{(n)}$ recursion relation (\ref{Arec}) we get
\begin{align}
\left[\delta A^{(n+1)}_\zeta + \right. & \left. \delta A^{(n)}_\zeta - \delta \tilde{A}^{(n+1)}_\zeta - \delta \tilde{A}^{(n)}_\zeta\right]^{(n)} =
[i\Lambda,A^{(n)}_\zeta] - [i\bar\Lambda, \tilde A^{(n)}_\zeta]\;\;.
\end{align}
After projecting on positive and negative powers of $\zeta$ we find what we need to prove the recursion relation. Then the full $A_\zeta$ transforms as
\begin{align}
\delta A_\zeta &= \sum_{n=1}^{\infty} \delta A^{(n)}_\zeta = \sum_{n=1}^{\infty} [\delta A^{(n)}_\zeta]^{(n-1)} +\sum_{n=1}^{\infty} [\delta A^{(n)}_\zeta]^{(n)}=
[\delta A^{(1)}_\zeta]^{(0)} + \sum_{n=1}^{\infty} [\delta A^{(n+1)}_\zeta
+\delta A^{(n)}_\zeta]^{(n)}\nn\\
&=-i\partial_\zeta\Lambda + [i\Lambda,A_\zeta]\;\;,
\end{align}
showing the correct transformation properties.

\section{A proof of the equivalence of the actions}
We are now in position to show that the standard action agrees with the one-loop calculation derived in the appendix.
\subsection{Equations of motion}
An indirect way of showing this would be to show that the one-loop action (\ref{action}) leads to the correct equations of motion
\begin{align}\label{fieldeq}
\bar{\nabla}^2\bar{\cal W} = 0 \;\;.
\end{align}
Starting from the variation of the action (\ref{action})
\begin{align}
\delta S = \sum_{n=2}^{\infty} (-1)^n\int d^8\theta \oint d\zeta_1\ldots d\zeta_n
\frac{\tr(\delta X_1 X_2\ldots X_n)}{\zeta_{21}\ldots \zeta_{n,n-1}\zeta_{1n}}\;\;,
\end{align}
we use the identity (\ref{dzeta}) on the $\frac1{\zeta_{1n}}$ factor to write
\begin{align}
\delta S &= \sum_{n=2}^{\infty} (-1)^n\int d^8\theta \oint d\zeta_1\ldots d\zeta_n
\frac{\tr(\delta X_1 X_2\ldots X_n)}{\zeta_{21}\ldots \zeta_{n,n-1}}\left(\delta_{n1} - \frac1{\zeta_{n1}}\right) =
\nn  \\ &
 \sum_{n=2}^{\infty} (-1)^n\int d^8\theta \oint d\zeta_1\ldots d\zeta_{n-1}
\frac{\tr(\delta X_1 X_2\ldots X_{n-1} X_1)}{\zeta_{21}\ldots \zeta_{1,n-1}}
\nn \\ &+\sum_{n=2}^{\infty}\int d^8\theta \oint d\zeta_1
\tr(\delta X_1 A^{(n-1)}_1)\;\;,
\end{align}
where we introduced the shorthand notation $A_i \equiv A_\zeta(\zeta_i)$.
Repeating the the last operation $(n-1)$ times we find
\begin{align}
\delta S = \sum_{n=2}^{\infty}\int d^8\theta \oint d\zeta_1
\sum_{k=1}^{n-1} (-1)^{k+1}\tr(\delta X_1 A_1^{(n-k)} X_1^{k-1}) +\nn \\
\sum_{n=2}^{\infty}(-1)^n\int d^8\theta \oint d\zeta_1 d\zeta_2 \frac{\tr(\delta X_1 X_1^{n-1})}{\zeta_{21}}\delta_{12}
\;\;,
\end{align}
where the last term, although singular when doing the remaining $\zeta$ integrals, actually vanishes when doing the $d^8\theta$ integral since it is projective.

It is possible to switch the order of the sums
\begin{align}
\delta S &= \sum_{k=1}^{\infty}\sum_{n=k+1}^{\infty}\int d^8\theta \oint d\zeta_1
 (-1)^{k+1}\tr(\delta X_1 A_1^{(n-k)} X_1^{k-1}) = \nn \\
 & \sum_{k=0}^{\infty}\sum_{n=1}^{\infty}\int d^8\theta \oint d\zeta_1
 (-1)^{k}\tr(\delta X_1 A_1^{(n)} X_1^{k}) =
 \int d^8\theta \oint d\zeta_1\tr(\delta X_1 A_1 (1+X_1)^{-1})\;\;.
\end{align}
Since $(1+X)^{-1} = e^{-V}$ we get the final result
\begin{align}\label{ActVar}
\delta S = \int d^8\theta \oint d\zeta_1\tr( e^{-V_1}\delta e^{V_1} A_1)\;\;.
\end{align}
This is in fact precisely what we need since in distinction to the variation $\delta V$, the expression $e^{-V}\delta e^V$ is gauge covariant and leads to covariant field equations. However, it is still constrained since $V$ is a projective superfield. We know that a projective superfield can be written in terms of a potential $X$ as $\nabla^2\bar\nabla^2 X$ and thus using integration by parts we derive the covariant field equations (\ref{fieldeq}).

\subsection{A direct proof}
The Yang-Mills action is given by (\ref{eq:YMA}). It is not obvious how it is related to the one-loop action (\ref{action}) although we know this has to be the case. We will now present a direct proof of this fact.
The ${\cal N}=2$ Yang-Mills theory action is
\begin{align}\label{ActWW}
S = & \frac12 \int [d^4\bar\theta]_C \tr(\bar{\mathbb W} \bar{\mathbb W}) = \frac12 \int [d^4\bar\theta]_C \int d\zeta_0
\frac{\tr(\bar{\cal W}_0\bar{\cal W}_0)}{\zeta_0} =\nn\\ &-\frac12
 \int [d^4\bar\theta]_C \int d\zeta_0 \frac{\tr(\nabla^2_0A_0\nabla^2_0 A_0)}{\zeta_0} =
  \frac14 \int [d^4\bar\theta]_C \int d\zeta_0 \nabla_0^2\frac{\tr(\nabla^\alpha_0A_0\nabla_{0\alpha} A_0)}{\zeta_0}
= \nn\\
&\frac14 \sum_{n=2}^{\infty}\sum_{k=1}^{n-1}\int [d^4\bar\theta]_C \int d\zeta_0 \nabla_0^2\frac{\tr(\nabla^\alpha_0A^{(k)}_0\nabla_{0\alpha} A^{(n-k)}_0)}{\zeta_0}\;\;,
\end{align}
where we have written the action as an expansion in terms of powers of $X=(e^V-1)$. To show that this is the same action as the action (\ref{action}) that we got from the one-loop calculation, we will reduce both to a convenient form.

We start with (\ref{ActWW}). Apart from the explicit $\frac{1}{\zeta_0}$ factor there are only positive powers of $\zeta_0$ in the action. Thus the $\zeta_0$ integral can be trivially performed with the result
\begin{align}
\frac14 \sum_{n=2}^{\infty}(-1)^n\sum_{k=1}^{n-1}\int [d^4\bar\theta]_C \oint d\zeta_1 \ldots d\zeta_n
D^2\tr\left(\frac{D^\alpha(X_1\ldots X_k)}{\zeta_1 \zeta_{21}\ldots\zeta_{k,k-1}\zeta_k}\;
\frac{D_\alpha(X_{k+1}\ldots X_n)}
{\zeta_{k+1}\zeta_{k+2,k+1}\ldots\zeta_{n,n-1}\zeta_n}\right),
\end{align}
where we have also expressed $A^{(n)}$ in terms of $X$ using (\ref{AsolX}). Acting with the $D$ derivatives and collecting terms of the same type (that is where the two supercovariant $D$ derivatives act on the same two $X$) we find that terms proportional to $\tr(X_1\ldots D^\alpha X_k \ldots D_\alpha X_{m+k-1}\ldots X_n)$ come with a $\zeta$ dependence of
\begin{align}
\frac{1}{\zeta_{21}\ldots\zeta_{n,n-1}\zeta_{1n}} \frac{(\zeta_1-\zeta_n)}{\zeta_1\zeta_n}
\sum_{r=k}^{m+k-2}\frac{(\zeta_{r+1}-\zeta_{r})}{\zeta_{r+1}\zeta_r} =
\frac{1}{\zeta_{21}\ldots\zeta_{n,n-1}\zeta_{1n}} \frac{(\zeta_1-\zeta_n)}{\zeta_1\zeta_n}
\frac{(\zeta_{m+k-1}-\zeta_{k})}{\zeta_{m+k-1}\zeta_{k}}\;\;.
\end{align}
When we finally use the cyclicity of the trace and relabel the $\zeta$ coordinates such that the $D^{\alpha}$ always act on $X_1$ and sum over all terms of the same type we reach the final form
\begin{align}\label{comp}
S = -\frac14 \sum_{n=2}^{\infty}(-1)^n\sum_{m=2}^{n}\int [d^4\bar\theta]_C \oint d\zeta_1 \ldots d\zeta_n
D^2\frac{\tr(D^\alpha X_1\ldots D_\alpha X_{m}\ldots X_n))}
{\zeta_{21}\ldots\zeta_{n,n-1}\zeta_{1n}} \frac{(\zeta_1-\zeta_m)^2}{\zeta_1^2\zeta_m^2}\;\;,
\end{align}
which is the final expression we would like to use for comparison.

If we on the other hand start with the one-loop action (\ref{action}) we begin by pushing in part of the fermionic measure and using the cyclicity of the trace to write
\begin{align}
\sum_{n=2}^{\infty}&\frac{(-1)^n}{n}\int [d^4\bar\theta]_C \oint d\zeta_1\ldots d\zeta_n D^2 Q^2\frac{\tr(X_1\ldots X_n)}{\zeta_{21}\ldots\zeta_{1n}}= \\
& \sum_{n=2}^{\infty} (-1)^n\int [d^4\bar\theta]_C \oint d\zeta_1\ldots d\zeta_n \frac{D^2}{\zeta_{21}\ldots\zeta_{1n}}\left[
\tr(Q^2X_1\ldots X_n)\right. \nn \\
&+ \left. \sum_{m=2}^{n} \frac{n-m+1}{n}\tr(Q^\alpha X_1\ldots Q_\alpha X_m\ldots X_n)
 \right] \;\;.\nn
\end{align}
We convert all $Q$ derivatives to $D$ derivatives using that the $X$ fields are projective
\begin{align}
\sum_{n=2}^{\infty} (-1)^n\int [d^4\bar\theta]_C \oint d\zeta_1\ldots d\zeta_n \frac{D^2}{\zeta_{21}\ldots\zeta_{1n}}\left[
\frac{1}{\zeta_1^2}\tr(D^2X_1\ldots X_n) \nn\right. \\ \left.
+ \sum_{m=2}^{n} \frac{n-m+1}{n}\frac{1}{\zeta_1\zeta_m}\tr(D^\alpha X_1\ldots D_\alpha X_m\ldots X_n)
 \right]\;\;.
\end{align}
In the first term we integrate one of the $D^\alpha$ by parts
\begin{align}
\sum_{n=2}^{\infty} & (-1)^n\int [d^4\bar\theta]_C \oint d\zeta_1\ldots d\zeta_n
\\
& \frac{D^2}{\zeta_{21}\ldots\zeta_{1n}}
\sum_{m=2}^{n} \frac{2(n-m+1)\zeta_1-n\zeta_m}{2n}\frac{1}{\zeta_1^2\zeta_m}\tr(D^\alpha X_1\ldots D_\alpha X_m\ldots X_n)\;\;.\nn
\end{align}
Two different terms where the $D_\alpha$ derivative acts on $X_m$ or $X_{m^\prime}$ respectively, in the case when $m+m^\prime = n+2$, can be made of the same type by using the cyclicity of the trace and cyclic relabeling of the $\zeta$ coordinates. Using this identification to "symmetrize" the coefficients we get
\begin{align}
-\frac14 \sum_{n=2}^{\infty} (-1)^n\sum_{m=2}^{n}\int [d^4\bar\theta]_C \oint d\zeta_1\ldots d\zeta_n \frac{D^2}{\zeta_{21}\ldots\zeta_{1n}}
\frac{(\zeta_1-\zeta_m)^2}{\zeta_1^2\zeta_m^2}\tr(D^\alpha X_1\ldots D_\alpha X_m\ldots X_n)\;\;,
\end{align}
which completely agrees with the expression (\ref{comp}) derived from (\ref{ActWW}).

\section{The Abelian limit}
A simple consistency check on the result is to go to the abelian limit and compare with the known form \cite{GonzalezRey:1997db}.
In the case where the gauge group is Abelian the recursion relation (\ref{Arec}) can be written as
\begin{align}
A^{(1)}_\zeta - \tilde{A}^{(1)}_\zeta &= \partial_\zeta X\;\;,\\
A^{(n+1)}_\zeta - \tilde{A}^{(n+1)}_\zeta &= -X(A^{(n)}_\zeta - \tilde{A}^{(n)}_\zeta)\;\;n\geq 1 \;\;,\nn
\end{align}
with the simple solution
\begin{align}
A^{(n+1)}_\zeta - \tilde{A}^{(n+1)}_\zeta = (-1)^{n} X^n \partial_\zeta X\;\;,
\end{align}
which directly sums to
\begin{align}
A_\zeta - \tilde{A}_\zeta = (1+X)^{-1}\partial_\zeta X = \partial_\zeta \ln (1+X) = \partial_\zeta V\;\;,
\end{align}
where we used that $X = e^V - 1$.

Starting from the action (\ref{action}) and using relation (\ref{ActVar}) we find
\begin{align}
S &= \sum_n \frac{(-1)^n}{n}\int d^8\theta \oint d\zeta_1\ldots d\zeta_n \frac{(X_1\ldots X_n)}{\zeta_{21}\ldots \zeta_{1n}} =
\nn \\ &
\int [d^4\bar\theta]_C \oint d\zeta_1 \frac12 D^2 Q^\alpha ( A_1 \left(e^{-V} Q_\alpha e^{V}\right)_1 ) = \nn \\
& \int [d^4\bar\theta]_C \oint d\zeta_1 d\zeta_2 \frac12 D^2 Q^\alpha
\frac{ V_2 Q_\alpha V_1}{\zeta_{21}^2} = \\
& \int [d^4\bar\theta]_C \oint d\zeta_1 d\zeta_2 \frac12 D^2 Q^\alpha
\frac{ V_2 Q_\alpha V_1}{\zeta_{21}}(\delta_{12}-\frac1{\zeta_{12}}) = \nn\\
&\int [d^4\bar\theta]_C \oint d\zeta_1 d\zeta_2 \frac12 D^2 Q^2\frac{V_1^2}{\zeta_{21}}\delta_{12}
-\int [d^4\bar\theta]_C \oint d\zeta_1 d\zeta_2 \frac12 D^2 Q^\alpha\frac{V_2Q_\alpha V_1}{\zeta_{21}\zeta_{12}}\;\;.
\nn
\end{align}
The first term vanishes since one can write $4D^2Q^2 = \Delta_1^2\nabla_1^2$ which gives zero when acting on the projective field $V_1^2$. The second term can be symmetrized in $\zeta_1$ and $\zeta_2$ to give
\begin{align}
-\frac14 \int [d^4\bar\theta]_C \oint d\zeta_1 d\zeta_2
 D^2 Q^\alpha\frac{V_2Q_\alpha V_1+V_1Q_\alpha V_2}{\zeta_{21}\zeta_{12}}\;\;,
\end{align}
which gives
\begin{align}
- \frac12 \int [d^4\bar\theta]_C \oint d\zeta_1 d\zeta_2
 D^2 Q^2\frac{V_1V_2}{\zeta_{21}\zeta_{12}}\;\;,
\end{align}
which agrees with the known action in the abelian case \cite{GonzalezRey:1997db} .

\section{Discussion}
We have found the kinetic term and interaction vertices for ${\cal N} = 2$ supersymmetric Yang-Mills theory in Projective superspace written in terms of the prepotential. Many of these results were already known but derived in a hybrid formalism between Projective and Harmonic superspace \cite{Jain:2009aj}. Here all calculations have been performed purely in Projective superspace. We have also derived several new results including the construction of the connection coefficient $A_\zeta$ (and thus the field strength $\bar{\cal W} = -i\nabla^2 A_\zeta$) in terms of $V$. As a consistency check we showed that our solution transforms correctly as a connection. We then used our solution in a direct proof that the action constructed from the divergent part of a one-loop calculation with a hypermultiplet running in the loop is equivalent to the action constructed using the field strength. As a check we showed that the action reduces to the known form in the abelian limit, something which is not at all obvious without using our solution.

Although the proofs in this paper were quite technical in nature the results themselves are simple and straightforward. We are convinced that the results and the techniques developed to prove them will be useful in the further development of gauge theory in Projective superspace. More concretely, with these new results it is now possible to continue with the quantization of the theory. We will continue the work started in \cite{GonzalezRey:1997db} for the abelian theory by investigating the gauge fixing and the derivation of the propagator and interaction terms for the full nonabelian theory.

We believe it to be of great importance to further develop Yang-Mills theory in Projective superspace as a complimentary tool for manifest $\mathcal{N}=2$ calculations. The $\epsilon$ prescription introduced in \cite{Jain:2009aj} gives an unambiguous method for how to deal with the $\mathbb{CP}^1$ integrals as compared with the treacherous coinciding harmonic singularities encountered in Harmonic superspace \cite{Buchbinder:1998np}. The methods developed in this paper will allow us to independently check results derived in Harmonic superspace where it is possible that problems of the type described above have led to erroneous results. We will also be able to proceed in situations where the Harmonic formalism is too complicated and where Projective superspace offers a simpler description. One such example is given by superconformal models of tensor multiplets where the Projective superspace formalism is clearly superior.
The results of this paper will also be necessary for studying gauge theories with eight supercharges in other dimensions. We are planning to apply our formalism to the six dimensional case as well as the case of $(4,p)$ supersymmetry in two dimensions.

\begin{acknowledgments}
RvU would like to thank Daniel Butter, Dharmesh Jain, Sergei Kuzenko, Ulf Lindstr\"{o}m, Martin Ro\v{c}ek, Warren Siegel and in particular Francisco Gonzalez-Rey for many enlightening discussions on the topic of this paper over the years.
This work is supported by the Grant agency of the Czech republic under the grant P201/12/G028.
\end{acknowledgments}

\appendix

\section{The kinetic term from a one-loop calculation}
One way to identify the kinetic term of the theory is to compute the contribution to the Yang-Mills effective action from a matter multiplet. The divergent part of the one-loop contribution has to be gauge invariant and proportional to the kinetic term for the result to be absorbed in a renormalization of the gauge coupling constant.

In projective superspace the interaction between a vector multiplet and a hypermultiplet
is through the term
\begin{align}
\int [d^4\theta]_P \frac{d\zeta}{\zeta} \bar{\Upsilon}e^V\Upsilon =
\int [d^4\theta]_P \frac{d\zeta}{\zeta} \left(\bar\Upsilon\Upsilon + \bar{\Upsilon}\left(e^V-1\right)\Upsilon\right)\;\;.
\end{align}
This gives rise to an interaction vertex
\begin{center}
\includegraphics[scale=0.6]{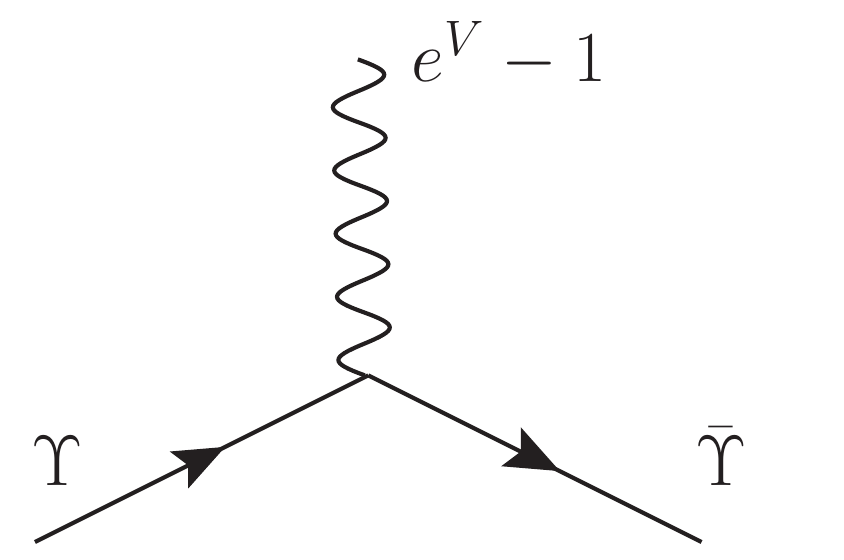}
\end{center}
where we need to remember that the vertex comes with a $[d^4 \theta]_P$ integral.

The polar propagator is \cite{GonzalezRey:1997qh}
\begin{align}
<\bar{\Upsilon}(1)\Upsilon(2)> = -\frac{1}{\zeta_1^2}\sum_{n=0}^{\infty}
\left(\frac{\zeta_2}{\zeta_1}\right)^n
\frac{\nabla_1^4\nabla_2^4}{(\zeta_1-\zeta_2)^2\Box}
\delta^4(x_1-x_2)\delta^8(\theta_1-\theta_2)\;\;,
\\
<\Upsilon(1)\bar{\Upsilon}(2)> = -\frac{1}{\zeta_2^2}\sum_{n=0}^{\infty}
\left(\frac{\zeta_1}{\zeta_2}\right)^n
\frac{\nabla_1^4\nabla_2^4}{(\zeta_1-\zeta_2)^2\Box}
\delta^4(x_1-x_2)\delta^8(\theta_1-\theta_2)\;\;,
\end{align}
where we will denote the direction of a propagator with an arrow.
Using this we can compute the one loop Feynman diagram with a hypermultiplet in the loop and factors of $(e^V-1)$ as
external lines.
\begin{center}
\includegraphics[scale=0.5]{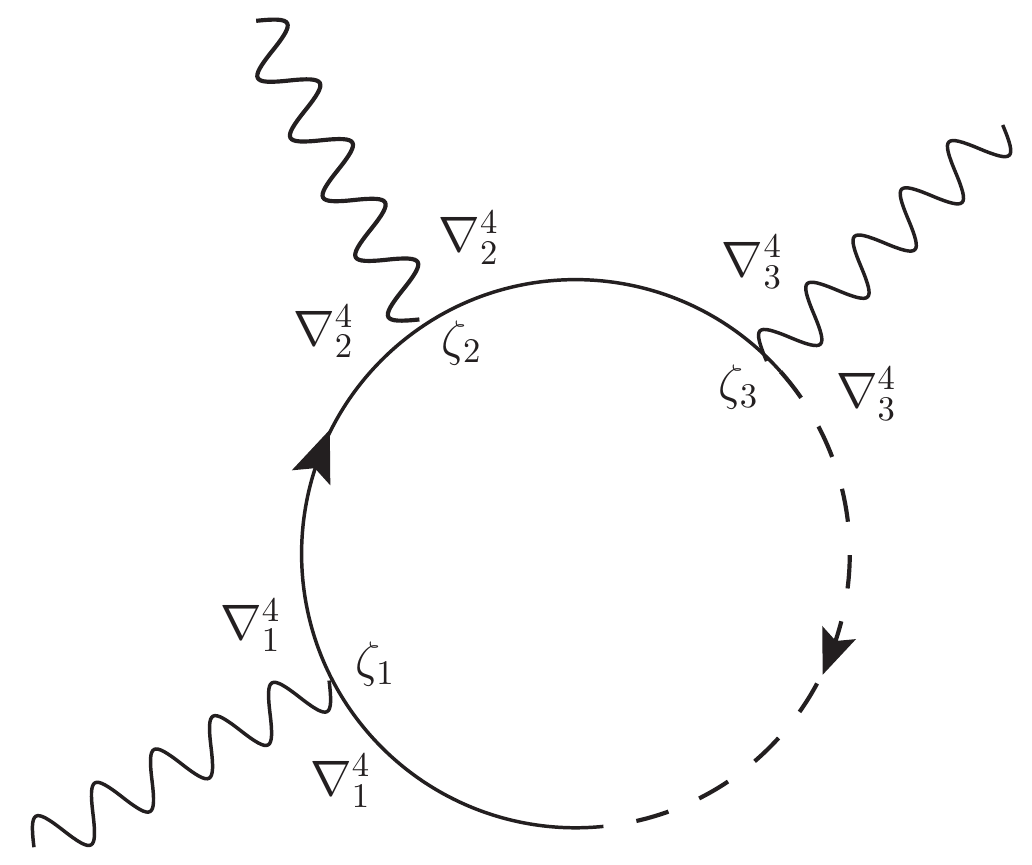}
\end{center}
For a diagram with $k$ vertices we begin by performing the D-algebra. First we need to complete the $[d^4\theta]_P$ measure of each vertex into a full superspace measure. This is done by absorbing a $\nabla^4$ from the propagator.
Using that $ [d^4\theta]_P \nabla^4 = d^8\theta \zeta^2$. The $\zeta^2$ factor will cancel the
$\frac{1}{\zeta^2}$ in each propagator.

If we are looking for the most divergent contribution, as many $\nabla$ derivatives as possible have to stay inside the loop
to cancel as many $\frac{1}{\Box}$ operators as possible. Using the identity
\begin{align}
\delta^8(\theta_{12})\nabla^4_1\ldots\nabla^4_k \delta^8(\theta_{21}) =
\Box^{k-2}(\zeta_1-\zeta_2)^2\ldots(\zeta_k - \zeta_1)^2 \delta^8(\theta_1-\theta_2) \;\;,
\end{align}
all the $(\zeta_i - \zeta_{i+1})^2$ cancel against the corresponding factors in the propagators.
What is left is
\begin{align}
\frac1{\Box^2}\frac{(-1)^n}{n}
\int d^8\theta \frac{d\zeta_1}{\zeta_1}\ldots
\frac{d\zeta_n}{\zeta_n} \sum_{k_1=0}^{\infty} \left(\frac{\zeta_1}{\zeta_2}\right)^{k_1}
\ldots \sum_{k_n=0}^{\infty} \left(\frac{\zeta_n}{\zeta_1}\right)^{k_n}
\tr\left(\left(e^V-1\right)_1\ldots \left(e^V-1\right)_n\right),
\end{align}
where the index $k$ in $(e^V-1)_k$ indicates that the superfield depends on $\zeta_k$ (all the $x$ and $\theta$ dependence is the same).

In the case that $|\zeta_1|<|\zeta_2|$ we can explicitly do the first sum
\begin{align}
\frac1{\zeta_2}\sum_{k_1=0}^{\infty}\left(\frac{\zeta_1}{\zeta_2}\right)^{k_1} = \frac1{\zeta_2-\zeta_1}\;\;.
\end{align}
Assuming that $|\zeta_i|<|\zeta_{i+1}|$ is always fulfilled we may continue to do the sums until we get to the final one where we have to assume that $|\zeta_n|<|\zeta_1|$ which is clearly impossible.
In order to write all the sums as factors of the type $\frac{1}{\zeta_{i+1}-\zeta_{i}}$ Jain and Siegel invented a prescription to perturb the $\zeta$-coordinates in such a way that the factor "remembers" from which sum it came \cite{Jain:2009aj}. To do this we think of all $\zeta$-coordinates to be at the same distance from the origin (take the unit circle for simplicity). Then we introduce an infinitesimal parameter $\epsilon$ that separates the $\zeta$-coordinates in the required way. We define
\begin{align}
\frac1{\zeta_{12}} \equiv& \frac1{\zeta_1-\zeta_2+\epsilon(\zeta_1+\zeta_2)} =
\frac1{(1+\epsilon)\zeta_1-(1-\epsilon)\zeta_2} = \nn\\
&\frac1{(1+\epsilon)\zeta_1} \sum_{n=0}^{\infty}
\left(\frac{(1-\epsilon)\zeta_2}{(1+\epsilon)\zeta_1}\right)^n \rightarrow
\frac1{\zeta_1}\sum_{n=0}^\infty \left(\frac{\zeta_2}{\zeta_1}\right)^n\;\;, \\
\frac1{\zeta_{21}} \equiv& \frac1{\zeta_2-\zeta_1+\epsilon(\zeta_2+\zeta_1)} =
\frac1{(1+\epsilon)\zeta_2-(1-\epsilon)\zeta_1} = \nn\\
&\frac1{(1+\epsilon)\zeta_2} \sum_{n=0}^{\infty}
\left(\frac{(1-\epsilon)\zeta_1}{(1+\epsilon)\zeta_2}\right)^n \rightarrow
\frac1{\zeta_2}\sum_{n=0}^\infty \left(\frac{\zeta_1}{\zeta_2}\right)^n \;\;.\nn
\end{align}
This allows us to keep track of from which type of sum the factors of $\frac1{\zeta_{i+1}-\zeta_i}$ came from which turns out to be essential for the calculations. A more thorough discussion of the $\eps$ prescription can be found in appendix C.

Using the $\epsilon$ prescription we may write the contribution concisely as
\begin{align}
\frac1{\Box^2}\frac{(-1)^n}{n}
\int d^8\theta \frac{d\zeta_1\ldots d\zeta_n}{\zeta_{21}\ldots\zeta_{1n}}
\tr\left(\left(e^V-1\right)_1\ldots \left(e^V-1\right)_n\right) \;\;.
\end{align}
All the fields are at the same $x$ and $\theta$ coordinate but at different
$\zeta$ coordinates as indicated by the index.

The momentum space integration gives an ultraviolet divergence through the $\frac1{\Box^2}$ term and we draw the conclusion that the action of the gauge field can be written as
\begin{align}\label{action}
S = \sum_{n=2}^{\infty} \frac{(-1)^n}{n}
\int d^8\theta \frac{d\zeta_1\ldots d\zeta_n}{\zeta_{21}\ldots\zeta_{1n}}
\tr\left(\left(e^V-1\right)_1\ldots \left(e^V-1\right)_n\right)\;\;.
\end{align}
This result was first given in \cite{Jain:2009aj}.

We may equivalently begin with
\begin{align}
\int [d^4\theta]_P \frac{d\zeta}{\zeta} {\Upsilon}e^{-V} \bar\Upsilon \;\;,
\end{align}
where $e^{-V}$ takes values in the complex conjugate representation (where the generators are $-T_A^\star$).
Then the propagators will have the $\zeta$-coordinates in the opposite order so the result becomes
\begin{align}
S = \sum_{n=2}^{\infty}\frac{(-1)^n}{n}
\int d^8\theta \frac{d\zeta_1\ldots d\zeta_n}{\zeta_{12}\ldots\zeta_{n1}}
\tr\left(\left(e^{-V}-1\right)_1\ldots \left(e^{-V}-1\right)_n\right)\;\;.
\end{align}

\section{Gauge invariance of the action}
As a further check on the result we may show that the action computed through the one-loop calculation is gauge invariant. We will see that the $\eps$ prescription is essential in deriving this result.

Starting from the action (\ref{action}) where $(e^V-1)$ transforms under infinitesimal gauge transformations as\footnote{This section reviews results that originally appeared in \cite{Jain:2009aj}}
\begin{align}
\delta(e^V-1) = i\bar\Lambda-i\Lambda+i\bar{\Lambda}\left(e^V-1\right)-\left(e^V-1\right)i\Lambda \;\;.
\end{align}
Focusing on the term with $n$ factors of $(e^V-1)$ and using the cyclicity of the trace as well as relabeling the $\zeta$-coordinates we can write the gauge transformation as
\begin{align}\label{eq:fullvar}
(-1)^n {\rm Tr} &\prod_{i=1}^{n}\oint d\zeta_i \frac{(i\bar\Lambda_1-i{\Lambda}_1)
(e^{V}-1)_2\cdot\ldots\cdot(e^{V}-1)_n}
{\zeta_{21}\cdot\ldots\cdot\zeta_{1n}}+
\nonumber\\
(-1)^n {\rm Tr} &\prod_{i=1}^{n}\oint d\zeta_i
\frac{(i\bar{\Lambda}_1 (e^{V}-1)_1-(e^{V}-1)_1i\Lambda_1)
(e^{V}-1)_2\cdot\ldots\cdot(e^{V}-1)_n}
{\zeta_{21}\cdot\ldots\cdot\zeta_{1n}}\;\;.
\end{align}
Starting with the first term we perform the $\zeta_1$ contour integral using the $\epsilon$ prescription. Since
\begin{align}\label{eq:var1}
-\frac{i\Lambda_1(e^{V}-1)_2\cdot\ldots\cdot(e^{V}-1)_n}
{\zeta_{21}\cdot\ldots\cdot\zeta_{n,n-1}}\;\;,
\end{align}
contains only positive powers of $\zeta_1$ (as dictated by the $\epsilon$ prescription),
we can do the $\oint d\zeta_1\frac{1}{\zeta_{1n}}$ integral which will just replace
$\zeta_1$ with $\zeta_n$ in (\ref{eq:var1}). Similarly, the term
\begin{align}
\frac{i\bar{\Lambda}_1(e^{V}-1)_2\cdot\ldots\cdot(e^{V}-1)_n}
{\zeta_{32}\cdot\ldots\cdot\zeta_{1n}}\;\;,
\end{align}
contains only negative powers of $\zeta_1$ so we can do the
$\oint d\zeta_1 \frac{1}{\zeta_{21}}$ integral which will replace $\zeta_1$
with $\zeta_2$ in the rest of the expression.
After using the cyclicity of the trace and relabeling the $\zeta$'s the full contribution of the first term in
(\ref{eq:fullvar}) is
\begin{align}
(-1)^{n-1} {\rm Tr}  \prod_{i=1}^{n-1} \oint d\zeta_i
\frac{((e^{V}-1)_1i\Lambda_1-i\bar{\Lambda}_1(e^{V}-1)_1)(e^{V}-1)_2
\cdot\ldots\cdot(e^{V}-1)_{n-1}}
{\zeta_{21}\cdot\ldots\cdot\zeta_{1,n-1}} \;\;.
\end{align}
This has the form of the second term in (\ref{eq:fullvar}) but with the opposite sign and one less power of $(e^V-1)$ so these two contributions will always cancel.

The only piece which does not cancel through this mechanism comes from the transformation of the
quadratic term
\begin{align}
{\rm Tr} \int d^8\theta \oint d\zeta_1 d\zeta_2 \frac{(i\bar\Lambda_1-i{\Lambda}_1)
(e^{V}-1)_2}{\zeta_{21}\zeta_{12}}\;\;.
\end{align}
We show that this vanish on each term separately. For instance, using (\ref{dzeta}) on the $\Lambda$ term we get
\begin{align}
{\rm Tr} \int d^8\theta \oint d\zeta_1 d\zeta_2 \frac{i{\Lambda}_1(e^{V}-1)_2}{\zeta_{21}\zeta_{12}}
= {\rm Tr} \int d^8\theta \oint d\zeta_1 d\zeta_2 \frac{i\Lambda_1(e^{V}-1)_2}{\zeta_{21}}
\left(\delta_{12} - \frac1{\zeta_{21}}\right)
\;\;.
\end{align}
The first term looks singular when doing the $\zeta$ integrals but vanishes since the integrand can be written as a projective superfield at a single $\zeta$ coordinate. The second term vanishes when doing the $\zeta_1$ integral since it contains only positive powers of $\zeta_1$. There is an analog argument for the term containing $\bar\Lambda$.

\section{The $\epsilon$ prescription}
The $\epsilon$ prescription was introduced in \cite{Jain:2009aj} as a way of remembering where certain factors of
$\frac{1}{\zeta_{12}}$ actually came from. The notation is the following
\begin{align}\label{epsdef}
 \frac{1}{\zeta_1}\sum_{n=0}^{\infty}
\left(\frac{\zeta_2}{\zeta_1}\right)^n \rightarrow \frac{1}{\zeta_{12}}\;\;.
\end{align}
Conversely we may define
\begin{align}
\frac1{\zeta_{12}} = \frac1{\zeta_1-\zeta_2+\eps(\zeta_1+\zeta_2)}\;\;.
\end{align}
When $\frac1{\zeta_{12}}$ appears in a contour integral where all contours are taken at the same distance from the origin $|\zeta_1| = |\zeta_2|$, the $\eps$ scales the $\zeta_1\rightarrow (1+\eps)\zeta_1$ and $\zeta_2 \rightarrow (1-\eps)\zeta_2$ so that $\frac1{\zeta_{12}}$ has to be expanded as in (\ref{epsdef}) and not the other way around. In a contour integral the $\frac1{\zeta_{12}}$ factors effectively function as projection operators on the positive or negative $\zeta$ powers in a projective superfield.
For example, if $X=\sum_{n=-\infty}^{\infty} X_n\zeta^n$ is a tropical superfield,
we may use the above definition to show that
\begin{align}
\sum_{n=0}^{\infty} X_n\zeta_2^n = \oint d\zeta_1 \frac{X(\zeta_1)}
{\zeta_{12}} \;\;,
\nonumber\\
\sum_{n=-\infty}^{-1} X_n\zeta_2^n = \oint d\zeta_1 \frac{X(\zeta_1)}
{\zeta_{21}}\;\;.
\end{align}

There are two relations that we are using repeatedly in this paper
\begin{align}
\delta_{12} &= \frac1{\zeta_{12}} + \frac1{\zeta_{21}} \label{dzeta}\;\;,\\
\frac1{\zeta_{12}\zeta_{23}} &= \frac1{\zeta_{13}\zeta_{23}} + \frac1{\zeta_{12}\zeta_{13}}\label{tri}\;\;,
\end{align}
where we again have suppressed a factor $2\pi i$ in the delta function corresponding to the suppressed $\frac1{2\pi i}$ in the measure. The relations are most easily proven by inserting them in a contour integral.

To illustrate how the $\eps$ prescription works we calculate a simple example. From the example we also learn that the $\frac1{\zeta_{12}}$ factors depending non-trivially on $\eps$ can be simply cancelled by non $\eps$ dependent factors of $(\zeta_1-\zeta_2)$ in the numerator.
In this appendix we will indicate the $\zeta$ dependence by two integer upper indices so that
\begin{align}
T^{(p,q)} = \sum_{k=p}^{q} T_k \zeta^k\;\;,
\end{align}
whereas a lower integer index will denote the particular coefficient in the $\zeta$ expansion.
In general we have
\begin{align}\label{IntUseP}
\oint d\zeta_1 \frac{1}{\zeta_{12}} T^{(0,\infty)}\left(\zeta_1\right) &= \frac1{1+\epsilon}T^{(0,\infty)}\left(\zeta_2\frac{1-\eps}{1+\eps}\right)\;\;, \\ \label{IntUseM}
\oint d\zeta_1 \frac{1}{\zeta_{21}} T^{(-\infty,-1)}\left(\zeta_1\right) &=
\frac1{1-\epsilon}T^{(-\infty,-1)}\left(\zeta_2\frac{1+\eps}{1-\eps}\right)\;\;.
\end{align}
Let $X(\zeta)$ and $Y(\zeta)$ be superfields with an arbitrary $\zeta$ dependence. Consider a term
\begin{align}
\oint d\zeta_1 d\zeta_2 \frac{1}{\zeta_{12}\zeta_{21}} (\zeta_1-\zeta_2)^2X(\zeta_1)Y(\zeta_2)\;\;.
\end{align}
Let us do the $\zeta_1$ integral. In order to use the relations (\ref{IntUseP}) and (\ref{IntUseM}) we first divide the integrand as
\begin{align}
(\zeta_1-\zeta_2)^2X(\zeta_1) &= \left[ \zeta_1^2 (X^{(-\infty,-3)}(\zeta_1) +X^{(-2,\infty)}(\zeta_1))
-2\zeta_1\zeta_2(X^{(-\infty,-2)}(\zeta_1)+X^{(-1,\infty)}(\zeta_1)) \right. \nn \\ &\left.
+\zeta_2^2(X^{(-\infty,-1)}(\zeta_1) + X^{(0,\infty)}(\zeta_1)) \right]\;\;.
\end{align}
The term with only positive powers (including the constant term) of $\zeta_1$ is
\begin{align}
T^{(0,\infty)} =
\frac1{\zeta_{21}}\left[
\zeta_1^2X^{(-2,\infty)}\left(\zeta_1\right)-2\zeta_1\zeta_2 X^{(-1,\infty)}\left(\zeta_1\right)
+ \zeta_2^2 X^{(0,\infty)}\left(\zeta_1\right)
\right]\;\;,
\end{align}
and what is left contains only negative powers
\begin{align}
T^{(-\infty,-2)} = \frac{1}{\zeta_{12}}\left[
\zeta_1^2X^{(-\infty,-3)}\left(\zeta_1\right) - 2\zeta_1\zeta_2 X^{(-\infty,-2)}\left(\zeta_1\right)
+ \zeta_2^2 X^{(-\infty,-1)}\left(\zeta_1\right)
\right]\;\;.
\end{align}
Using (\ref{IntUseP}) and (\ref{IntUseM}) we do the integral over $\zeta_1$. In particular we have
\begin{align}
\frac{1}{\zeta_{21}}  &\rightarrow
\frac{1}{\zeta_2-\zeta_2\frac{1-\eps}{1+\eps} + \eps(\zeta_2 + \zeta_2\frac{1-\eps}{1+\eps})} =
\frac{1+\eps}{4\eps\zeta_2}\;\;,\\
\frac{1}{\zeta_{12}}  &\rightarrow
\frac{1}{\zeta_2\frac{1+\eps}{1-\eps}-\zeta_2 + \eps(\zeta_2 + \zeta_2\frac{1+\eps}{1-\eps})} =
\frac{1-\eps}{4\eps\zeta_2}\;\;,
\end{align}
so that the integral of the term with positive powers give
\begin{align}
\frac{1}{4\eps\zeta_2}\left[
\left(\zeta_2\frac{1-\eps}{1+\eps}\right)^2 \right. & X^{(-2,\infty)}\left(\zeta_2\frac{1-\eps}{1+\eps}\right) \nn \\ & \left.
 -2\left(\zeta_2\frac{1-\eps}{1+\eps}\right)\zeta_2 X^{(-1,\infty)}\left(\zeta_2\frac{1-\eps}{1+\eps}\right) \right. \nn \\ & \left.
 +\zeta_2^2 X^{(0,\infty)}\left(\zeta_2\frac{1-\eps}{1+\eps}\right)
\right] = \nn \\  =
\frac{\zeta_2}{4\eps} \left[
\left(\frac{2\eps}{1+\eps}\right)^2 \right. & \left. X^{(0,\infty)}\left(\zeta_2\frac{1-\eps}{1+\eps}\right)
+X_{-2}\frac{1}{\zeta_2^2} - \frac{1+3\eps}{1+\eps}X_{-1}\frac{1}{\zeta_2}
\right]\;\;,
\end{align}
and for the negative powers we have
\begin{align}
\frac{1}{4\eps\zeta_2}\left[
\left(\zeta_2\frac{1+\eps}{1-\eps}\right)^2 \right. & X^{(-\infty,-3)}\left(\zeta_2\frac{1+\eps}{1-\eps}\right) \nn \\
& -2\left(\zeta_2\frac{1+\eps}{1-\eps}\right)\zeta_2 X^{(-\infty,-2)}\left(\zeta_2\frac{1+\eps}{1-\eps}\right) \nn \\ &\left.
+\zeta_2^2 X^{(-\infty,-1)}\left(\zeta_2\frac{1+\eps}{1-\eps}\right)
\right]  = \nn \\ =
\frac{\zeta_2}{4\eps} \left[
\left(\frac{2\eps}{1-\eps}\right)^2 \right. & \left. X^{(-\infty,-3)}\left(\zeta_2\frac{1+\eps}{1-\eps}\right)
-\frac{1+2\eps-3\eps^2}{(1+\eps)^2} X_{-2}\frac{1}{\zeta_2^2} + \frac{1-\eps}{1+\eps}X_{-1}\frac{1}{\zeta_2}
\right]\;\;.
\end{align}
Adding things up we get
\begin{align}
\frac{\zeta_2}{4\eps}\left[
\left(\frac{2\eps}{1-\eps}\right)^2 X^{(-\infty,-3)}\left(\zeta_2\frac{1-\eps}{1+\eps}\right)+
\left(\frac{2\eps}{1+\eps}\right)^2 X^{(0,\infty)}\left(\zeta_2\frac{1-\eps}{1+\eps}\right) \nonumber \right. \\ \left.
+\left(1-\frac{1+2\eps-3\eps^2}{(1+\eps)^2} \right)X_{-2}\frac{1}{\zeta_2^2}
+\left(\frac{1-\eps}{1+\eps}- \frac{1+3\eps}{1+\eps}\right) X_{-1}\frac{1}{\zeta_2}
\right] = \nn\\
\frac{\zeta_2}{4\eps}\left[
\left(\frac{2\eps}{1-\eps}\right)^2 X^{(-\infty,-3)}\left(\zeta_2\frac{1-\eps}{1+\eps}\right)+
\left(\frac{2\eps}{1+\eps}\right)^2 X^{(0,\infty)}\left(\zeta_2\frac{1-\eps}{1+\eps}\right) \nonumber \right. \\ \left.
+\left(\frac{2\eps}{1+\eps}\right)^2 X_{-2}\frac{1}{\zeta_2^2}
- \frac{4\eps}{1+\eps} X_{-1}\frac{1}{\zeta_2}
\right]\;\;.
\end{align}
We find that all inverse powers of $\eps$ cancel and the limit $\eps\rightarrow 0$ is smooth
\begin{align}
\oint d\zeta_1 d\zeta_2 \frac{1}{\zeta_{12}\zeta_{21}} (\zeta_1-\zeta_2)^2X(\zeta_1)Y(\zeta_2)=
-\oint d\zeta_2 X_{-1} Y(\zeta_2) = -X_{-1}Y_{-1}\;\;,
\end{align}
which is the same result we would have gotten if we naively would have cancelled the $(\zeta_1-\zeta_2)^2$ factor against the $\frac{1}{\zeta_{12}\zeta_{21}}$ denominator.


\begin{thebibliography}{123}
\bibitem{Karlhede:1984vr}
  A.~Karlhede, U.~Lindstr\"om and M.~Ro\v{c}ek,
  ``Selfinteracting Tensor Multiplets in $N=2$ Superspace,''
  Phys.\ Lett.\  {\bf 147B} (1984) 297.
  doi:10.1016/0370-2693(84)90120-5


\bibitem{Hitchin:1986ea}
  N.~J.~Hitchin, A.~Karlhede, U.~Lindstr\"om and M.~Ro\v{c}ek,
  ``Hyperkahler Metrics and Supersymmetry,''
  Commun.\ Math.\ Phys.\  {\bf 108} (1987) 535.
  doi:10.1007/BF01214418


\bibitem{Karlhede:1986mg}
  A.~Karlhede, U.~Lindstr\"om and M.~Ro\v{c}ek,
  ``Hyperkahler Manifolds and Nonlinear Supermultiplets,''
  Commun.\ Math.\ Phys.\  {\bf 108} (1987) 529.
  doi:10.1007/BF01214417


\bibitem{Lindstrom:2008gs}
  U.~Lindstr\"om and M.~Ro\v{c}ek,
  ``Properties of hyperkahler manifolds and their twistor spaces,''
  Commun.\ Math.\ Phys.\  {\bf 293} (2010) 257
  doi:10.1007/s00220-009-0923-0
  [arXiv:0807.1366 [hep-th]].


\bibitem{deWit:2001brd}
  B.~de Wit, M.~Ro\v{c}ek and S.~Vandoren,
  ``Hypermultiplets, hyperKahler cones and quaternion Kahler geometry,''
  JHEP {\bf 0102} (2001) 039
  doi:10.1088/1126-6708/2001/02/039
  [hep-th/0101161].


\bibitem{Gaiotto:2008cd}
  D.~Gaiotto, G.~W.~Moore and A.~Neitzke,
  ``Four-dimensional wall-crossing via three-dimensional field theory,''
  Commun.\ Math.\ Phys.\  {\bf 299} (2010) 163
  doi:10.1007/s00220-010-1071-2
  [arXiv:0807.4723 [hep-th]].

\bibitem{Crichigno:2012vd}
  P.~M.~Crichigno and D.~Jain,
  ``Darboux Coordinates and Instanton Corrections in Projective Superspace,''
  JHEP {\bf 1210} (2012) 027
  doi:10.1007/JHEP10(2012)027
  [arXiv:1204.3899 [hep-th]].


\bibitem{Lindstrom:1989ne}
  U.~Lindstr\"om and M.~Ro\v{c}ek,
  ``$N=2$ Superyang-mills Theory in Projective Superspace,''
  Commun.\ Math.\ Phys.\  {\bf 128} (1990) 191.
  doi:10.1007/BF02097052


\bibitem{GonzalezRey:1997db}
  F.~Gonzalez-Rey,
  ``Feynman rules in N=2 projective superspace. 3. Yang-Mills multiplet,''
  hep-th/9712128.


\bibitem{GonzalezRey:1997qh}
  F.~Gonzalez-Rey, U.~Lindstr\"om, M.~Ro\v{c}ek, R.~von Unge and S.~Wiles,
  ``Feynman rules in N=2 projective superspace: 1. Massless hypermultiplets,''
  Nucl.\ Phys.\ B {\bf 516} (1998) 426
  doi:10.1016/S0550-3213(98)00073-X
  [hep-th/9710250].


\bibitem{GonzalezRey:1997xp}
  F.~Gonzalez-Rey and R.~von Unge,
  ``Feynman rules in N=2 projective superspace. 2. Massive hypermultiplets,''
  Nucl.\ Phys.\ B {\bf 516} (1998) 449
  doi:10.1016/S0550-3213(98)00074-1
  [hep-th/9711135].


\bibitem{Kuzenko:2007cj}
  S.~M.~Kuzenko and G.~Tartaglino-Mazzucchelli,
  ``Five-dimensional Superfield Supergravity,''
  Phys.\ Lett.\ B {\bf 661} (2008) 42
  doi:10.1016/j.physletb.2008.01.055
  [arXiv:0710.3440 [hep-th]].


\bibitem{Kuzenko:2007hu}
  S.~M.~Kuzenko and G.~Tartaglino-Mazzucchelli,
  ``5D Supergravity and Projective Superspace,''
  JHEP {\bf 0802} (2008) 004
  doi:10.1088/1126-6708/2008/02/004
  [arXiv:0712.3102 [hep-th]].


\bibitem{Kuzenko:2008wr}
  S.~M.~Kuzenko and G.~Tartaglino-Mazzucchelli,
  ``Super-Weyl invariance in 5D supergravity,''
  JHEP {\bf 0804} (2008) 032
  doi:10.1088/1126-6708/2008/04/032
  [arXiv:0802.3953 [hep-th]].


\bibitem{Kuzenko:2008ep}
  S.~M.~Kuzenko, U.~Lindstr\"om, M.~Ro\v{c}ek and G.~Tartaglino-Mazzucchelli,
  ``4D N = 2 Supergravity and Projective Superspace,''
  JHEP {\bf 0809} (2008) 051
  doi:10.1088/1126-6708/2008/09/051
  [arXiv:0805.4683 [hep-th]].


\bibitem{Kuzenko:2009zu}
  S.~M.~Kuzenko, U.~Lindstr\"om, M.~Ro\v{c}ek and G.~Tartaglino-Mazzucchelli,
  ``On conformal supergravity and projective superspace,''
  JHEP {\bf 0908} (2009) 023
  doi:10.1088/1126-6708/2009/08/023
  [arXiv:0905.0063 [hep-th]].


\bibitem{Kuzenko:2010bd}
  S.~M.~Kuzenko,
  ``Lectures on nonlinear sigma-models in projective superspace,''
  J.\ Phys.\ A {\bf 43} (2010) 443001
  doi:10.1088/1751-8113/43/44/443001
  [arXiv:1004.0880 [hep-th]].


\bibitem{Galperin:1984av}
  A.~Galperin, E.~Ivanov, S.~Kalitsyn, V.~Ogievetsky and E.~Sokatchev,
  ``Unconstrained N=2 Matter, Yang-Mills and Supergravity Theories in Harmonic Superspace,''
  Class.\ Quant.\ Grav.\  {\bf 1} (1984) 469
   Erratum: [Class.\ Quant.\ Grav.\  {\bf 2} (1985) 127].
  doi:10.1088/0264-9381/1/5/004


\bibitem{Zupnik:1986ca}
  B.~M.~Zupnik,
  ``Solution of Constraints of Supergauge Theory in the Harmonic SU(2)/U(1) Superspace,''
  Theor.\ Math.\ Phys.\  {\bf 69} (1986) 1101
   [Teor.\ Mat.\ Fiz.\  {\bf 69} (1986) 207].
  doi:10.1007/BF01037868


\bibitem{Zupnik:1987vm}
  B.~M.~Zupnik,
  ``The Action of the Supersymmetric $N=2$ Gauge Theory in Harmonic Superspace,''
  Phys.\ Lett.\ B {\bf 183} (1987) 175.
  doi:10.1016/0370-2693(87)90433-3


\bibitem{Galperin:2007wpa}
  A.~S.~Galperin, E.~A.~Ivanov, V.~I.~Ogievetsky and E.~S.~Sokatchev,
  ``Harmonic Superspace,''

\bibitem{rosly1983super}
A.~A.~Rosly, ``Super Yang-Mills  constraints as integrability conditions,'' in Proceedings of the International Seminar
{\it Group Theoretical Methods in Physics} (Zvenigorod, USSR, 1982), M. A. Markov  (Ed.), Nauka, Moscow, 1983, Vol. 1, p. 263 (in Russian);
English translation: in {\it Group Theoretical Methods in Physics},'' M. A. Markov, V. I. Man'ko  and A. E. Shabad  (Eds.), Harwood Academic Publishers,
London, Vol. 3, 1987, p. 587.


\bibitem{Kuzenko:1998xm}
  S.~M.~Kuzenko,
  ``Projective superspace as a double punctured harmonic superspace,''
  Int.\ J.\ Mod.\ Phys.\ A {\bf 14} (1999) 1737
  doi:10.1142/S0217751X99000889
  [hep-th/9806147].

\bibitem{Butter:2012ta}
  D.~Butter,
  ``Relating harmonic and projective descriptions of N=2 nonlinear sigma models,''
  JHEP {\bf 1211} (2012) 120
  doi:10.1007/JHEP11(2012)120
  [arXiv:1206.3939 [hep-th]].

\bibitem{Jain:2009aj}
  D.~Jain and W.~Siegel,
  ``Deriving Projective Hyperspace from Harmonic,''
  Phys.\ Rev.\ D {\bf 80} (2009) 045024
  doi:10.1103/PhysRevD.80.045024
  [arXiv:0903.3588 [hep-th]].


\bibitem{Jain:2010gm}
  D.~Jain and W.~Siegel,
  ``On Projective Hoops: Loops in Hyperspace,''
  Phys.\ Rev.\ D {\bf 83} (2011) 105024
  doi:10.1103/PhysRevD.83.105024
  [arXiv:1012.3758 [hep-th]].


\bibitem{Jain:2012jx}
  D.~Jain and W.~Siegel,
  ``A Note on Massive Scalar Hypermultiplet in Projective Hyperspace,''
  Phys.\ Rev.\ D {\bf 86} (2012) 065036
  doi:10.1103/PhysRevD.86.065036
  [arXiv:1106.4601 [hep-th]].


\bibitem{Jain:2012zx}
  D.~Jain and W.~Siegel,
  ``N=2 Super-Yang-Mills Theory from a Chern-Simons Action,''
  Phys.\ Rev.\ D {\bf 86} (2012) 125017
  doi:10.1103/PhysRevD.86.125017
  [arXiv:1203.2929 [hep-th]].


\bibitem{Jain:2013hua}
  D.~Jain and W.~Siegel,
  ``Improved Methods for Hypergraphs,''
  Phys.\ Rev.\ D {\bf 88} (2013) no.2,  025018
  doi:10.1103/PhysRevD.88.025018
  [arXiv:1302.3277 [hep-th]].


\bibitem{Kuzenko:2014mva}
  S.~M.~Kuzenko and G.~Tartaglino-Mazzucchelli,
  ``N = 4 supersymmetric Yang-Mills theories in $AdS_3$,''
  JHEP {\bf 1405} (2014) 018
  doi:10.1007/JHEP05(2014)018
  [arXiv:1402.3961 [hep-th]].


\bibitem{Butter:2014xxa}
  D.~Butter, S.~M.~Kuzenko, J.~Novak and G.~Tartaglino-Mazzucchelli,
  ``Conformal supergravity in five dimensions: New approach and applications,''
  JHEP {\bf 1502} (2015) 111
  doi:10.1007/JHEP02(2015)111
  [arXiv:1410.8682 [hep-th]].

\bibitem{Buchbinder:1998np}
  I.~L.~Buchbinder and S.~M.~Kuzenko,
  ``Comments on the background field method in harmonic superspace: Nonholomorphic corrections in N=4 SYM,''
  Mod.\ Phys.\ Lett.\ A {\bf 13} (1998) 1623
  doi:10.1142/S0217732398001704
  [hep-th/9804168].

\end{thebibliography}
\end{document}